\documentclass[12pt]{article}
\usepackage{amscd,amsthm}
\usepackage{latexsym}
\usepackage{amsmath}
\usepackage{amssymb}
\numberwithin{equation}{section}
\def\bbZ{{\Bbb Z}}
\def\bbR{{\Bbb R}}

\def\Re{{\rm Re}}
\def\Im{{\rm Im}}
\def\const{{\rm const}}
\def\Trace{{\rm Trace}}
\def\var{{\rm Var}}

\def\a{{\cal A}}

\def\f{{\cal F}}

\def\k{{\cal K}}
\def\n{{\cal N}}

\def\p{{\cal P}}

\begin{document}
\title{Gaussian Fluctuation for the Number of Particles in Airy, Bessel, Sine
and Other Determinantal Random Point Fields}
\author{Alexander B. Soshnikov\\
California Institute of Technology\\Department of
Mathematics\\Sloan 253-37\\Pasadena, CA  91125, USA\\
and\\
University of California, Davis\\
Department of Mathematics,\\ One Shields Ave., Davis, CA 95616, USA}
\date{December 1999}
\maketitle
\begin{abstract}
We prove the Central Limit Theorem for the number of eigenvalues near the spectrum
edge for hermitian ensembles of random matrices.  To derive our results, we
use a general theorem, essentially due to Costin and Lebowitz, 
concerning the Gaussian fluctuation of the number of particles in random
point fields with determinantal correlation functions.  
As another corollary of Costin-Lebowitz Theorem we prove CLT for the 
empirical distribution function of the eigenvalues of random matrices from
classical compact groups.
\end{abstract}

\section{Introduction and Formulation of Results}

Random hermitian matrices were introduced in mathematical physics by Wigner
in the fifties ([Wig1], [Wig2]).  The main motivation of pioneers in this 
field was 
to obtain a better understanding of the statistical behavior of energy
levels of heavy nuclei.  An archetypical example of random matrices is the 
Gaussian Unitary Ensemble (G.U.E.) which can be defined by the probability
distribution on a space of $n$-dimensional hermitian matrices as
\begin{equation}
P(dA)=\const_n\cdot e^{-2n\ \Trace\ A^2}dA.
\end{equation}
Here $dA$ is the Lebesgue measure on $n^2$-parameters set 
\begin{equation}
\{\Re\ a_{ij},
1\leq i<j\leq n;\ \Im\ a_{ij},\ 1\leq i<j\leq n,\ a_{ii},\ 1\leq i\leq n\}
\end{equation}
and const$_n=(\pi n)^{-\frac{n^2}{2}}\cdot 2^{n(2n-1)/2}$ is a
normalization constant.  (1.1) implies that matrix entries (1.2) are 
independent Gaussian random variables $N(0,\frac{1+\delta_{ij}}{8n})$.
It is well known that the G.U.E. is the only ensemble of hermitian
random matrices (up to a trivial rescaling) that satisfies both of the
following properties:
\begin{enumerate}
\item[(1)] probability distribution $P(dA)$ is invariant under unitary
transformation
$$A\rightarrow U^{-1}AU,\ U\in U(n),$$

\item[(2)] matrix entries up from the diagonal are independent random
variables (see [Me], Ch. 2).
\end{enumerate}
The $n$ eigenvalues, all real, of hermitian matrix $A$ will be denoted by
$\lambda_1,\break\lambda_2,\dots ,\lambda_n$.  For the formulas for their
joint distribution density $p_n(\lambda_1,\ldots ,\lambda_n)$ and 
$k$-point correlation functions $\rho_{n,k}(\lambda_1,\ldots ,\lambda_k)$ we
refer to Mehta's book [Me].  One has
\begin{equation}
p_n(\lambda_1,\ldots ,\lambda_n)=\const '_n\cdot\prod_{1\leq i<j\leq n}\vert
\lambda_i-\lambda_j\vert^2\cdot exp (-2n\cdot\sum^n_{i=1}\lambda^2_i)
\end{equation}
\begin{align}
\begin{split}
\rho_{n,k}(\lambda_1,\ldots ,\lambda_k):&=\frac{n!}{(n-k)!}\int_{\bbR^{n-k}}
p_n(\lambda_1,\ldots ,\lambda_n)d\lambda_{k+1}\ldots d\lambda_n\\
&=\det (K_n(\lambda_i,\lambda_j))^k_{i,j=1},
\end{split}
\end{align}
where $K_n(x,y)$ is a projection kernel,
\begin{equation}
K_n(x,y)=\sqrt{2n}\cdot\sum^{n-1}_{\ell =0}\psi_\ell (\sqrt{2n}x)\cdot
\psi_\ell (\sqrt{2n}\cdot y)
\end{equation}
and
\begin{equation}
\psi_\ell (x)=\frac{(-1)^\ell}{\pi^{\frac{1}{4}}\cdot (2^\ell\cdot\ell 
!)^{\frac{1}{2}}}\cdot \exp\left (\frac{x^2}{2}\right )\cdot\frac{d^\ell}
{dx^\ell}\exp (-x^2)
\end{equation}
$\ell =0,1,\dots$, are Weber-Hermite functions.
The global behavior of eigenvalues is governed by 
the celebrated semicircle law,
which states that the empirical distribution function of the eigenvalues
weakly converges to a non-random (Wigner) distribution:
\begin{equation}
\f_n(\lambda )=\frac{1}{n}\#\{\lambda_i\leq \lambda\}\overset{w}
{\underset{n\rightarrow\infty}{\longrightarrow}}\f(\lambda )=
\int^\lambda_{-\infty}\rho (x)dx
\end{equation}
with probability one ([Wig1], [Wig2]) where the spectral density $\rho$ is given by
\begin{equation}
\rho (t)=\begin{cases} \frac{2}{\pi}\sqrt{1-t^2}, & \vert t\vert\leq 1\\
0,& \vert t\vert >1.
\end{cases}
\end{equation}
To study the local behavior of eigenvalues near an arbitrary point in the 
spectrum $x\in [-1,1]$, one has to consider rescaling
\begin{equation}
\lambda_i=x+\frac{y_i}{\rho_{n,1}(x)},\ i=1,\ldots ,k
\end{equation}
and study the rescaled $k$-point correlation functions
\begin{equation}
R_{n.k}(y_1,\ldots ,y_k):=(\rho_{n,1}(x))^{-k}\cdot\rho_{n,k}(\lambda_1,
\dots ,\lambda_k).
\end{equation}
The biggest interest is paid to the asymptotics of rescaled correlation
functions
when $n$ goes to infinity.  For G.U.E. the answer can be obtained from the
Plancherel-Rotach asymptotic formulas for Hermite polynomials ([PR]):
\begin{equation}
\lim_{n\rightarrow\infty}R_{n,k}(y_1,\ldots ,y_k)=\rho_k(y_1,\ldots ,y_k)=
\det (K(y_i, y_j))^k_{i,j=1}.
\end{equation}
The  $K$ actually also depends on $ x$ but in a very simple way.  It can be represented as
\begin{equation}
K(y,z)=\frac{\a (y)\cdot \a '(z)-\a (z)\a '(y)}{y-z},
\end{equation}
where for all $\vert x\vert <1$ the function $\a$ is just $\frac{\sin 
(\pi y)}{\pi}$, and for $x=\pm 1$ it is
\begin{equation}
\a i(\pm y)=\frac{1}{\pi}\int^\infty_0\cos \left (\frac{1}{3}t^3\pm yt\right
)dt.
\end{equation}
The function defined by (1.13) is known as the Airy function and the kernel
(1.12)--(1.13) is known as the Airy kernel (see [Me], [TW1], [F]).  
The limiting
correlation functions (1.11)-(1.13) determine a random point field on the 
real line, i.e., probability measure on the Borel $\sigma$-algebra of the
space of locally finite configurations,
\begin{equation}
\Omega =\{\omega =(y_i)_{i\in\bbZ}:\forall T>0\ \#\{y_i:\vert y_i\vert <T\}
<\infty\}.
\end{equation}
The distribution of random point field is uniquely defined by the generating
function
$$\psi (z_1,\ldots ,z_k;\ I_1,\ldots ,I_k)=E\prod^k_{j=1}z^{\nu_j}_j,$$
where $I_j,\ j=1,\ldots ,k$, are disjoint intervals on the real line,
$\nu_j=\#\{y_i\in I_j\}=\#(I_j)$, the number of particles in $I_j$,
and $k\in\bbZ^1_+$.  It follows from the general theory of existence and
uniqueness for random point fields ([L1], [L2]), that if $K(y,z)$ is locally 
bounded than determinantal correlation functions
uniquelly determine random point field assuming that  such randompoint fiels exists.  The
generating function $\psi (z_1,\ldots ,z_k)$ is given by Fredholm determinant
of the intergal operator in $L^2(\bbR^1)$:
\begin{equation}
\psi (z_1,\ldots ,z_k)=\det (\delta (x-y)+\sum^k_{j=1} (z_j-1)\cdot
K(x,y)\cdot \chi_{I_j}(y)),
\end{equation}
where $\chi_{I_j}$ is an indicator of $I_j$.

In particular these results are applicable to the Airy kernel (1.12)--(1.13).
We shall call the corresponding random point field the Airy random point
field.  For one-level density formulas (1.11)--(1.13) produce
\begin{equation}
\rho_1(y)=-y\cdot(\a i)^2(y)+(\a i '(y))^2.
\end{equation}
The asymptotic expansion of the Airy function is well known (see [Ol]).
One can deduce from it
\begin{equation}
\rho_1(y)\sim\begin{cases}\frac{\vert y\vert^{\frac{1}{2}}}{\pi}-\frac{\cos
(4\cdot\vert y\vert^{\frac{3}{2}}/3)}{4\pi\cdot\vert y\vert}+\underline{0}
(\vert y\vert^{-\frac{5}{2}})\ \text{as }y\rightarrow -\infty,\\
\frac{17}{96\cdot\pi y^{\frac{1}{2}}}\cdot \exp (-4y^{\frac{3}{2}}/3)
\ \text{as }y\rightarrow +\infty .\end{cases}
\end{equation}
$\rho_1$ satisfies the third order differential equation
\begin{equation}
\rho_1'''(y)=-2\rho_1(y)+4y\cdot\rho_1'(y).
\end{equation}
One can think about the one-point correlation function as a level 
density, since
for any interval $I\subset\bbR^1$  we have $E\#(I)=\int_I\rho_1(y)dy$.
It follows from (1.17) that $E\#((-T,+\infty ))$ is finite for any $T$
and $E\#((-T,+\infty))\sim\frac{2T^{\frac{3}{2}}}{3\pi}+\underline{0}
(1)$ when $T$ goes to $+\infty$.  
The last formula means that $E\#((-T,+\infty ))-\frac{2T^{\frac{3}{2}}}
{3\pi}$ stays bounded for large positive $T$.  Let us denote
$\nu_1 (T):=\#\{y_i>-T\}=\#((-T,+\infty )), \ \ 
\nu_k (T):=\#((-kT, -(k-1)T] )), k=2,3, \ldots. $
Theorem 1
establishes the Central Limit Theorem for $\nu_k (T)$.
\medskip

\noindent{\bf Theorem 1}
{\it
The variance of $\nu_k (T)$ grows logarithmically 
$${\rm Var}\ \nu_k (T)\sim\frac{11}
{12\pi^2}\cdot\log T+\underline{0}(1),$$ 
and  the sequence of
normalized random variables
$\frac{\nu_k (T)-E\nu_k (T)}{\sqrt {Var\ \nu_k (T)}}$ converges in distribution
to the centalized gaussian random sequence $\{ \xi_k \}$ with the covariance 
function 
$E \xi_k \xi_l = \delta_{k,l} -1/2 \ \delta_{k,l+1}- 1/2 \ \delta_{k,l-1}$.}
\medskip

\noindent{\it Remark 1}.
The first result about Gaussian fluctuation of the number of particles
in random matrix model was established by Costin and Lebowitz
([CL]) for the kernel $\frac{\sin\pi (x-y)}{\pi (x-y)}$.  See \S 2 for a
more detailed discussion.

\noindent{\it Remark 2}.
Basor and Widom ([BaW]) recently proves the Central Limit
Theorem for a large class of smooth linear statistics 
$\sum^{+\infty}_{i=-\infty}f(y_i/T)$ where $f$ satisfies some decay and
differentiality conditions.  Similar results for smooth linear statistics
in other random matrix ensembles were proven in [Sp],[DS], [Jo1], [SiSo1], [KKP],
[SiSo2], [Ba], [BF],  [BdMK], [Br];
see also [So1]
for the results about global distribution of spacings.

Another class of random hermitian matrices, called Laguerre ensemble, was
introduced by Bronk in [Br].  This one is the ensemble of positive
$n\times n$ hermitian matrices.  Any positive hermitian matrix $H$ can be
represented as $H=AA^*$, where $A$ is some complex valued $n\times n$
matrix and $A^*$ is its conjugate.  The distribution on such matrices is
defined as
\begin{equation}
P(dH)=\const ''_n\cdot\exp (-n\cdot \Trace\ A\cdot A^*)\cdot [\det (AA^*)]^\alpha
dA,
\end{equation}
where $\alpha >-1$ and $dA$ is Lebesgue measure on $2n^2$-dimensional space
of complex matrices.  The joint distribution of $n$ (positive) eigenvalues
of $H$ is given by
\begin{equation}
p_n(\lambda_1,\ldots ,\lambda_n)=\const '''_n\cdot\exp\left
(-n\cdot\sum^n_{i=1}
\lambda_i\right )
\cdot\prod^n_{i=1}\lambda^\alpha_i\cdot\prod_{1\leq i<j\leq n}
(\lambda_i-\lambda_j)^2.
\end{equation}
The Vandermonde factor in (1.20) implies that correlation
functions still have the determinantal form (1.4) with the kernel
\begin{equation}
K_n(x,y)=n\cdot\sum^{n-1}_{\ell =0}\phi_\ell (nx)\cdot\phi_\ell (ny),
\end{equation}
where the sequence $\{\phi_\ell (x)\}$ is obtained by orthonormalizing the
sequence $\{x^k\cdot x^{\frac{\alpha}{2}}\cdot e^{-x/2}\}$ on $(0,+\infty )$.
The limiting level density is supported on $[0,1]$ and given by the equation
\begin{equation}
\rho_1(x)=\frac{1}{2\pi}x^{-\frac{1}{2}}\cdot (1-x)^{\frac{1}{2}}.
\end{equation}
(It is not surprising to see that (1.22) is the density of a square of the Wigner
random variable!)  Plancherel-Rotach type asymptotics for Laguerre
polynomials ([E], [PR]) imply that by scaling the kernel $K_n(x,y)$
in the bulk of the spectrum, we obtain the sine kernel $\frac{\sin\pi 
(x-y)}{\pi (x-y)}$, and by scaling at $x=1$ (``soft edge"), we obtain the Airy
kernel.  As we already know, the same kernels appear after rescaling in 
G.U.E.  This feature, called universality of local correlations, has
been established recently for a variety of ensembles of hermitian random
matrices (see [PS],[BI],[DKMVZ],[Jo2],[So2],[BZ]).  To scale the kernel 
at the ``hard edge" $x=0$,
we need an asymptotic formula of Hilb's type (see [E]), which leads to
\begin{equation}
\lim_{n\rightarrow\infty}\frac{1}{4n}K_n\left (\frac{x}{4n},\frac{y}{4n}
\right )=\frac{J_\alpha (\sqrt x)\cdot\sqrt y\cdot J'_\alpha (\sqrt y)
-\sqrt xJ'_\alpha (x)\cdot J_\alpha (y)}{2(x-y)},
\end{equation}
where $J_\alpha$ is the Bessel function of order $\alpha$ ([F], [TW2], [Ba]).
The kernel (1.23) is also known to appear at hard edges in 
the Jacobi ensemble
([NW]).  For a quick reference, we note that in the Jacobi case, a sequence
$\{\phi_\ell (x)\}$ from (1.21) is obtained by orthonormalizing
$\{x^k(1-x)^{\frac{\alpha}{2}}(1+x)^{\frac{\beta}{2}}\}$. 
The random point field
on $[0,+\infty ]$ with the determinantal correlation functions defined by (1.23)
will be referred to as the Bessel random point field.  
There is a general belief among people working in random
matrix theory that in the same way as the sine kernel appears to be 
a universal 
limit in the bulk of the spectrum for random hermitian matrices, Airy and
Bessel kernels are universal limits at the soft and hard edge of
the spectrum.  
The next theorem establishes
the CLT for $\nu_k (T)=\#(((k-1)T,kT]), k=1,2,\ldots $.
\medskip

\noindent{\bf Theorem 2}
{\it
Let $\nu (T)$ be the number of particles in $(0, T)$ for the
Bessel random point field.  Then
$$E\ \nu_k (T)\sim\frac{1}{\pi}T^{\frac{1}{2}}(k^{1/2}- (k-1)^{1/2}
+\underline{0}(1),$$
$${\rm Var}\ \nu_k (T)\sim\frac{1}{4\pi^2}\log T+\underline{0}(1),$$
and the sequence of the
the normalized random variable $\frac{\nu_k (T)-E\ \nu_k (T)}{\sqrt{Var\ \nu_k
(T)}}$ converges in distribution to the gaussian random sequence from the Theorem 1.}
\medskip

Theorems 1 and 2, as well as similar results for the random 
fields arising from
the classical compact groups (see [So1])
are the corollaries of the general
result about determinantal random point fields, which is essentially due to
Costin and Lebowitz. 
Recently a number of discrete determinantal random point fields appeared in 
two-dimensional growth models (
[Jo3],[Jo5]), asymptotics of Plancherel measures on symmetric groups 
and the representation theory
of the infinite symmetric group ([BO1], [BO2],[BOO],[Jo4],[Ok1],[Ok2]).
 If one can show the infinite growth 
of the 
variance of the number of particles in these models ( the goal which may be 
 probably attainable since the 
asymptotics of the discrete orthogonal polynomials arising in some of these problems 
are known) the Costin- Lebowitz theorem should work there as well.

The rest of the paper is organized as follows.  We discuss
the general (Costin-Lebowitz) theorem in \S 2.  Theorems
1 and 2 will be proven in \S 3 and \S 4.  In the Bessel case, we will see that
the kernel $\frac{\sin\pi (x-y)}{\pi (x-y)}\pm\frac{\sin\pi (x+y)}{\pi
(x+y)}$ naturally appears in our considerations.  We recall
in \S 4 that the sine kernel also appears in the limiting distribution
of eigenvalues in unitary group and the even 
and odd sine kernels  appear in the distribution of eigenvalues in 
orthogonal and symplectic groups and then prove Theorems 3-6 the Gaussian 
fluctuation for the
number of eigenvalues in these models in Theorem 3-6.

It is a pleasure to thank Prof. Ya. Sinai, who drew my attention to the
preprint of Costin-Lebowitz paper some time ago, and the organizers
of the Special Session on Integrable Systems and Random Matrix Theory (AMS
Meeting at Tucson, November 13--15, 1998) and the Introductory Workshop
on Random Matrix Models (MSRI, Berkeley, January 19--23, 1999) for the 
opportunity to attend the meetings, where the idea of this paper has been
finalized.
The work was partially supported by the Euler stipend from the German Mathematical
Society.

\section{The Central Limit Theorem for Determinantal Random Point Fields}

Let $\{\p_t\}_{t\in\bbR^1_+}$ be a family of random point fields on the 
real line such that their correlation functions have determinantal form
at the r.h.s. of
(1.11) with kernels $K_t(y,z)$, and $\{I_t\}_{t\in\bbR^1_+}$ a set
of intervals.  We denote by $A_t$ an integral operator
on $I_t$ with the kernel $K_t(y,z),\ A_t: L^2(I_t)\rightarrow L^2(I_t)$,
by $\nu_t$ the number of particles in $I_t,\ \nu_t=\# (I_t)$, and by
$E_t,\ {\rm Var}_t$ 
the mathematical expectation and variance with respect to the
probability distribution of the random field $\p_t$.  In many applications
the random point field $\p_t$, and therefore the
kernel $K_t$ will be the same for all $t$.  In such situations
the interval $I_t$ will be expanding.
\medskip

\noindent{\bf Theorem} (O. Costin, J. Lebowitz)
{\it
Let $A_t=K_t\cdot\chi_{I_t}$ be a family of
trace class  operators associated with determinantal random point fields
$\{\p_t\}$
such that 
{\rm Var}$_t\ \nu_t=\Trace
(A_t-A^2_t)$ goes to infinity as $t\rightarrow +\infty$. 
Then the distribution
of the normalized random variable $\frac{\nu_t-E_t\nu_t}{\sqrt{Var_t\ 
\nu_t}}$ with respect to the random point field
$\p_t$ weakly converges to the normal law $N(0,1)$.}
\medskip

\noindent{\it Remark 3}.
The result has been proven by Costin and Lebowitz when $K_t(x,y)=\frac
{\sin\pi (x-y)}{\pi (x-y)}$ for any $t$ and $\vert I_t\vert\underset
{t\rightarrow \infty}{\longrightarrow}\infty$ (see [CL]).  The original
paper contains a remark, due to Widom, that the result holds for more
general kernels.

\noindent{\it Remark 4}.
There is a general result that a (locally) trace class operator K
defines a determinantal random point field iff $ 0 \leq K \leq 1 \ \ $ (see [ So4] 
or, for a slightly weaker version, [Ma]).

The idea of the proof is very clear and consists of two parts.  Let us
denote the $\ell$th cumulant of $\nu_t$ by $C_\ell (\nu_t)$.
We remind that by definition
$$ \sum_{\ell=1}^{\infty} C_\ell(iz)^{\ell} / \ell ! = \log \left (E_t \exp 
( i z \nu_t) \right ) .$$
\medskip

\noindent{\bf Lemma 1}
{\it The following recursive relation holds for any $\ell\geq 2$:
\begin{equation}
C_\ell (\nu_t)=(-1)^\ell\cdot (\ell -1)!\Trace (A_t-A_t^\ell )+\sum^{\ell -1}
_{s=2}\alpha_{s\ell}C_s(\nu_t),
\end{equation}
where $\alpha_{s\ell},\ 2\leq s\leq \ell -1$, are some combinatorial 
coefficients (irrelevant for our purposes).}
\medskip

The proof can be found in [CL] or [So1], \S 2; (of course one has
to replace everywhere $\frac{\sin\pi (x-y)}{\pi (x-y)}$ by $K_t (x,y)$).
For the convinience of the reader we sketch the main ideas here. We start
by introducing the Ursell (cluster) functions :
$$
r_1(x_1)= \rho(x_1) , \ \ r_2(x_1,x_2)= \rho_2(x_1,x_2)- \rho_1(x_1) \rho(
x_2) ,$$
and, in general,
\begin{equation}
r_k(x_1, \ldots , x_k) = \sum_{m=1}^k \sum\limits_{G} (-1)^{m-1} (m-1)! \ 
 \prod_{j=1}^m \ \ r_{G_j}(\bar x(G_j) )
\end{equation}
where $  \ \ G \ \ $ is a partition of indices $ \ \  \{ 1,2,\ldots , k \} \ $
into $m$ subgroups $ \ G_1, \ldots G_m ,  \ \ $ and
$ \ \ \bar x(G_j) \ \ $ stands for the collection of $ \ x_i \ $ with 
indices in $ \ \ G_j \ \  .$ \\
It appears that the integral of k-point Ursell function $ \ \ r_k (x_1,
\ldots , x_k) \ \ $ over $k$-dimensional cube $ I_t \times \ldots I_t $
is equal to the linear combination of $ C_j(\nu_t), \ \ j=1,\ldots k .$.
Namely, let us denote
$$
T_k(\nu_t)= \int_{I_t} \ldots \int_{I_t} \ r_k(x_1, \ldots, x_k) dx_1
\ldots dx_k
$$
Then
\begin{equation}
 \sum_{k}^{\infty} C_k(iz)^{k} / k ! = \sum_{k=1}^{\infty}
( \exp (z) -1)^{k} T_{k}(\nu_t) / k ! 
\end{equation}
Taking into account that for the determinantal random point fields
$$
T_k(\nu_t)= (-1)^k \cdot (k-1)! Trace (A_t)^k,
$$
the last two equations imply (2.1).
The next lemma allows us to estimate Trace $(A_t-A^\ell_t)$.
\medskip

\noindent{\bf Lemma 2}
{\it 
$\ 0\leq\ \Trace (A_t-A^\ell_t)\leq (\ell -1)\cdot \Trace\ (A_t-A^2_t)$.}
\medskip

The proof is elementary:
$0\leq \Trace\ (A_t-A^\ell_t)=\sum^{\ell -1}_{j=1}\ \Trace\ (A^j_t-A^{j+1}_t)
\leq\sum^{\ell -1}_{j=1}\Vert A^{j-1}_t\Vert\cdot\ \Trace\ (A_t-A^2_t)\leq
(\ell -1)\cdot\ \Trace\ (A_t-A^2_t).$\qed

As a corollary of the lemmas we have $C_\ell (\nu_t)=\underline{0}(C_2
(\nu_t))$ for any $\ell\geq 2$. Since $C_2(\nu_t)=\Trace\ (A_t-A^2_t)
\underset{t\rightarrow\infty}{\longrightarrow}+\infty$, we conclude that
for $\ell >2$, $C_\ell (\frac{\nu_t-E\nu_t}{\sqrt{var_t\nu_t}})=\frac{C_\ell
(\nu_t)}{((C_2(\nu_t))^\ell /2}\underset{t\rightarrow\infty}{\longrightarrow}
0$.

At the same time the first two cumulants of the normalized random variable
are 0 and 1, respectively.  The convergence of cumulants implies the
convergence of moments to the moments of $N(0,1)$.  The theorem is proven.
\qed

To generalize the Costin-Lebowitz theorem to the case of several intervals we consider $ I_t^{(m)}, \ m+1, \ldots , s $, disjoint intervals of the 
real line, and define $ \nu_t^{(m)} = \# (I_t^{(m)}).$
The equation
\begin{equation}
\sum C_{k_1, \ldots , k_s}(iz_1)^{k_1} / k_1 ! \cdots (iz_s)^{k_s} /
k_s ! = \log \left (E_t \exp ( i (z_1 \nu_t^{(1)}+ \ldots + z_s \nu_t^{(s)})
 ) \right )
\end{equation}
defines the joint cumulants of $\nu_t^{(m)}$'s.
\medskip

\noindent{\bf Proposition 1}
{\it 
$C_{k_1, \ldots, k_s} \left (\nu_t^{(1)},\ldots, \nu_t{(s)}\right ) $ is equal 
to the linear combination of the traces
$$ \Trace K_t \cdot \chi_{I_t}^{(\cdots)} \cdot  K_t \cdot 
\chi_{I_t}^{(\cdots)} \ldots  K_t \cdot \chi_{I_t}^{(\cdots)} $$
with some combinatorial coefficients (irrelevant for our purposes), such
that for any $k_j$ that is greater than zero at least one indicator in each term of the linear combination is the indicator of $I_t^{(j)} .$ }
\medskip

The proof immedeately follows from the analogue of (2.4) for the case of
a several intervals.

In the next section we will apply theses results  to prove Theorem 1.

\section{Proof of Theorem 1}
For the most part of the section we will study the case of one interval
$ (-T, +\infty )$.
We start by recalling the asymptotic expansion of Airy function for large
positive and negative $y$ (see [Ol]).
\begin{equation}
\a_i(\vert y\vert )\sim\frac{e^{-\pi z}}{2\pi^{\frac{1}{2}}
\cdot\vert y\vert^{\frac{1}{4}}}\cdot\sum^\infty_{s=0}(-1)^s\cdot\frac
{u_s}{z^s};
\end{equation}
\begin{equation}
\a_i'(\vert y\vert )\sim\frac{\vert y\vert^{\frac{1}{4}}\cdot e^{-\pi z}}
{2\pi^{\frac{1}{2}}}\cdot\sum^\infty_{s=0}(-1)^s\cdot\frac{v_s}{z^s}
\end{equation}
\begin{align}
\begin{split}
&\a_i(-\vert y\vert )\sim\frac{1}{\pi^{\frac{1}{2}}\cdot\vert y
\vert^{\frac{1}{4}}}\cdot\biggl\lbrace\cos \left (\pi z+\frac{\pi}{4}
\right )\cdot
\sum^\infty_{s=0}(-1)^s\cdot\frac{u_{2s}}{z^{2s}}\\
&\qquad +\sin (\pi z+\frac{\pi}{4})\cdot\sum^\infty_{s=0}(-1)^s\cdot
\frac{u_{2s+1}}{z^{2s+1}}\biggr\rbrace ,
\end{split}
\end{align}
\begin{align}
\begin{split}
&\a_i' (-\vert y\vert )\sim\frac{\vert y\vert^{\frac{1}{4}}}
{\pi^{\frac{1}{2}}}\cdot\biggl\lbrace\sin \left (\pi z+\frac{\pi}{4}
\right )\cdot
\sum^\infty_{s=0}(-1)^{s+1}\cdot\frac{v_{2s}}{z^{2s}}\\
&\qquad -\cos (\pi z+\frac{\pi}{4})\cdot\sum^\infty_{s=0}(-1)^{s+1}\cdot
\frac{v_{2s+1}}{z^{2s+1}}\biggr\rbrace ,
\end{split}
\end{align}
where $z=\frac{2}{3\pi}y\cdot\vert y\vert^{\frac{1}{2}};\ u_0=v_0=1$, and
$u_s=\frac{(2s+1)\cdot (2s+3)\cdot\ldots\cdot (6s-1)}{(216\cdot\pi )^2
\cdot s!}$, $v_s=-\frac{6s+1}{6s-1}u_s,\ s\geq 1$.  In particular, as a
consequence of (3.1)--(3.4) one has (1.17).
It follows from (3.1)--(3.4) together with the boundedness of $\a_i(y)$,
$\a_i'(y)$ on any compact set that for any fixed $a\in\bbR^1$ all moments
of $\#((a,+\infty ))$ are finite.  Therefore it is enough to establish
the CLT for $\#((-T,a))$.  We choose  $a=-(\frac{3\pi}{2})^{\frac{2}{3}}\ 
(y=-(\frac{3\pi}{2})^{\frac{2}{3}}$ corresponds to $z=-1$).  We are going
to show that the conditions of the theorem from \S 2 are satisfied by $K\cdot
\chi_{(-T,a)}$, where as above, this notation is reserved for the integral
operator with the kernel $K(x,y)\cdot\chi_{(-T, a)}(y)$.
\medskip

\noindent{\bf Lemma 3}
{\it 
$\ 0\leq K\cdot\chi_{(-T,a)}\leq 1$ and $K\cdot\chi_{(-T,a)}$ is trace 
class.\qed}
\medskip

The kernel $K(y_1,y_2)=\frac{\a_i(y_1)\cdot\a)i'(y_2)-\a_i'
(y_1)\cdot\a_i(y_2)}{y_1-y_2}$ was obtained from \break
$K_n(x_1,x_2)=\sqrt{2n}\cdot\sum^{n-1}_{\ell =0}\psi_\ell (\sqrt{2n}x_1)\cdot
\psi_\ell (\sqrt{2n}\cdot x_2)$ after rescaling $x_i=1+\frac{y_i}{2n^{\frac
{2}{3}}}$, $i=1,2,$ and taking the limit $n\rightarrow\infty$.  The 
convergence  is uniform on compact sets.  As a projection operation
$K_n$ satisfies $0\leq K_n\leq 1$.  We immediately conclude that $K\cdot
\chi_{(-T,a)}$ satisfies the same inequalities and since the kernel is
continuous and non-negative definite the operator is trace class
(see e.g [GK] or [RS], section XI.4).  Now the main
step of the proof consists of
\medskip

\noindent{\bf Proposition 2}.{\it
$${\rm Var} \left (\#\left (-T, -\left ( \frac{3\pi}{2}\right )^{\frac{2}{3}}\right )\right )\sim\frac{11}{12\pi^2}
\log T+\underline{0}(1).$$}
\medskip

\noindent{\it Proof}.  We introduce the change of variables 
\begin{equation}
z_i=\frac{2}{3\pi}y_i\cdot\left\vert y_i\right\vert^{\frac{1}{2}}
\end{equation}
and agree to use the notations $Q(z_1,z_2),\ q_k(z_1,\dots ,z_k),\ k=1,2,
\dots ,$ for the kernel and $k$-point correlation function of the new
random point field obtained by (3.5).  It follows from (1.17) that
$q_1(z)\sim 1+\tfrac{\cos 2\pi z}{6\pi z}+\underline{0} (z^{-2})$ for
$z\rightarrow -\infty$, so we see that the configuration $(z_i)$ is
equally spaced at $-\infty$.

The kernel $Q(z_1,z_2)$ is defined by
\begin{align*}
\begin{split}
Q(z_1,z_2)&=\frac{\pi}{\vert y_1\vert^{\tfrac{1}{4}}\cdot\vert y_2
\vert^{\tfrac{1}{4}}}\cdot K(y_1,y_2)\\
&=\frac{\pi}{\vert y_1\vert^{\tfrac{1}{4}}\cdot\vert y_2
\vert^{\tfrac{1}{4}}}\cdot
\frac{\a_i(y_1)\cdot\a_i'(y_2)-\a '_i(y_1)\cdot\a_i(y_2)}{y_1-y_2}.
\end{split}
\end{align*}
Formulas (3.3)--(3.4) allow us to represent $Q$ as the sum of six
kernels $Q^{(i)},\ i=1,\dots ,6$, with the known asymptotic expansion:
$Q(z_1,z_2)=\sum^6_{i=1}Q^{(i)}(z_1,z_2)$, where

\begin{align}
\begin{split}
&Q^{(1)}(z_1,z_2)\sim\frac{1}{3\pi}\cdot\frac{1}{z_1^{\tfrac{2}{3}}
-z_2^{\tfrac{2}{3}}}\cdot\sin\pi (z_1-z_2)
\cdot \biggl\lbrace \sum^\infty_{m,n=0}(-1)^{m+n}\\
&\qquad \cdot u_{2m}\cdot v_{2n}
\cdot \left (z_1^{-2m-\tfrac{1}{3}}\cdot z_2^{-2n}+z_1^{-2n}\cdot 
z_2^{-2m-\tfrac{1}{3}}\right )\biggr\rbrace ;
\end{split}
\end{align}

\begin{align}
\begin{split}
&Q^{(2)}(z_1,z_2)\sim\frac{1}{3\pi}\cdot\frac{1}{z_1^{\tfrac{2}{3}}
-z_2^{\tfrac{2}{3}}}\cdot\cos\pi (z_1+z_2)\cdot \\
&\qquad \biggl\lbrace \sum^\infty_{m,n=0}(-1)^{m+n}
\cdot u_{2m}\cdot v_{2n}
\cdot \left (z_1^{-2n}\cdot z_2^{-2m-\tfrac{1}{3}}
-z_1^{-2m-\tfrac{1}{3}}\cdot z_2^{-2n}\right )\biggr\rbrace ;
\end{split}
\end{align}

\begin{align}
\begin{split}
&Q^{(3)}(z_1,z_2)\sim\frac{1}{3\pi}\cdot\frac{1}{z_1^{\tfrac{2}{3}}
-z_2^{\tfrac{2}{3}}}\cdot 2\cos\left (\pi z_1+\frac{\pi}{4}\right )
\cdot\cos\left (\pi z_2+\frac{\pi}{4}\right )
\cdot\\
&\qquad \biggl\lbrace \sum^\infty_{m,n=0}
(-1)^{m+n+1}
\cdot u_{2m}\cdot v_{2n+1}
\cdot \left (z_1^{-2m-\tfrac{1}{3}}\cdot z_2^{-2n-1}
-z_1^{-2n-1}\cdot z_2^{-2m-\tfrac{1}{3}}\right )\biggr\rbrace ;
\end{split}
\end{align}

\begin{align}
\begin{split}
&Q^{(4)}(z_1,z_2)\sim\frac{1}{3\pi}\cdot\frac{1}{z_1^{\tfrac{2}{3}}
-z_2^{\tfrac{2}{3}}}\cdot 2\sin\left (\pi z_1+\frac{\pi}{4}\right )
\cdot\sin\left (\pi z_2+\frac{\pi}{4}\right )
\cdot\\ 
&\qquad \biggl\lbrace\sum^\infty_{m,n=0} (-1)^{m+n}
\cdot u_{2m+1}\cdot v_{2n}
\cdot \left (z_1^{-2m-\tfrac{4}{3}}\cdot z_2^{-2n}
-z_1^{-2n}\cdot z_2^{-2m-\tfrac{4}{3}}\right )\biggr\rbrace ;
\end{split}
\end{align}

\begin{align}
\begin{split}
&Q^{(5)}(z_1,z_2)\sim\frac{1}{3\pi}\cdot\frac{1}{z_1^{\tfrac{2}{3}}
-z_2^{\tfrac{2}{3}}}\cdot\sin\pi (z_1-z_2) 
\cdot\\
&\qquad \biggl\lbrace \sum^\infty_{m,n=0} (-1)^{m+n+1}
\cdot u_{2m+1}\cdot v_{2n+1}
\cdot \left (z_1^{-2m-\tfrac{4}{3}}\cdot z_2^{-2n-1}
+z_1^{-2n-1}\cdot z_2^{-2m-\tfrac{4}{3}}\right )\biggr\rbrace ;
\end{split}
\end{align}

\begin{align}
\begin{split}
&Q^{(6)}(z_1,z_2)\sim\frac{1}{3\pi}\cdot\frac{1}{z_1^{\tfrac{2}{3}}
-z_2^{\tfrac{2}{3}}}\cdot\cos\pi (z_1+z_2)
\cdot\\
&\qquad  \biggl\lbrace \sum^\infty_{m,n=0}(-1)^{m+n}
\cdot u_{2m+1}\cdot v_{2n+1}
\cdot \left (z_1^{-2n-1}\cdot z_2^{-2m-\tfrac{4}{3}}
-z_1^{-2m-\tfrac{4}{3}}\cdot z_2^{-2n-1}\right )\biggr\rbrace .
\end{split}
\end{align}
We denote by $Q^{(i)}_{m,n}(z_1,z_2)$ the $(m,n)^{th}$ term in the
asymptotic expansion of $Q^{(i)}(z_1,z_2)$.  Then 
\begin{align*}
\begin{split}
Q^{(1)}_{0,0}
(z_1,z_2)&=\frac{\sin\pi (z_1-z_2)}{z_1^{\tfrac{2}{3}}-z_2^{\tfrac{2}{3}}}
\cdot\frac{1}{3\pi}\cdot\left (z_1^{-\tfrac{1}{3}}+z_2^{-\tfrac{1}{3}}
\right )\\
&=\frac
{\sin\pi (z_1-z_2)}{\pi (z_1-z_2)}\cdot\frac{z_1^{\tfrac{2}{3}}
+z_1^{\tfrac{1}{3}}\cdot z_2^{\tfrac{1}{3}}+z_2^{\tfrac{2}{3}}}
{3\cdot z_1^{\tfrac{1}{3}}\cdot z_2^{\tfrac{1}{3}}}.
\end{split}
\end{align*}
We note that near
the diagonal $Q^{(1)}_{0,0}$ is essentially a sine kernel.  We also will
need 
$$Q^{(2)}_{0,0}(z_1,z_2)=\frac{\cos\pi (z_1+z_2)}{\pi (z_1+z_2)}\cdot
\frac{z_1^{\tfrac{2}{3}}
-z_1^{\tfrac{1}{3}}\cdot z_2^{\tfrac{1}{3}}+z_2^{\tfrac{2}{3}}}
{3\cdot z_1^{\tfrac{1}{3}}\cdot z_2^{\tfrac{1}{3}}}.$$  
Let us define
$S(z_1,z_2)=Q^{(1)}_{0,0}(z_1,z_2)+Q^{(2)}_{0,0}(z_1,z_2),\ U(z_1,z_2)=
Q(z_1,z_2)-S(z_1,z_2)$.
\medskip

\noindent{\bf Lemma 4}.  {\it
\begin{equation}
\int^{-1}_{-L} \int^{-1}_{-L}\left (Q^{(1)}_{0,0}(z_1,z_2)\right )^2dz_1dz_2
=L-\frac{2}{3\pi^2}\log L+\underline{0}(1).
\end{equation}
}

\noindent{\it Proof}.  The integral can be written as
\begin{align*}
\begin{split}
&\frac{1}{9}\int^L_1\int^L_1\left (\frac{\sin\pi (z_1-z_2)}{\pi (z_1-z_2)}
\right )^2\cdot\left (
\frac{z_1^{\tfrac{2}{3}}+z_1^{\tfrac{1}{3}}\cdot z_2^{\tfrac{1}{3}}
+z_2^{\tfrac{2}{3}}}{z_1^{\tfrac{1}{3}}\cdot z_2^{\tfrac{1}{3}}}\right )^2
dz_1dz_2\\
&\quad =\frac{2}{9}\cdot\int^{L-1}_0\left (\frac{\sin\pi u}{\pi u}
\right )^2\cdot\int^{L-u}_1\left [\left (\frac{z+u}{z}\right )^{\frac{1}{3}}
+1+\left (\frac{z}{z+u}\right )^{\frac{1}{3}}\right ]^2dzdu.
\end{split}
\end{align*}
We represent the inner integral as
\begin{align}
\begin{split}
&\int^{L-u}_1\left [\left (\frac{z+u}{z}\right )^{\frac{2}{3}}+2\cdot
\left (\frac{z+u}{z}\right )^{\frac{1}{3}}+
\left (\frac{z}{z+u}\right )^{\frac{2}{3}}+2\cdot
\left (\frac{z}{z+u}\right )^{\frac{1}{3}}
+3\right ]dz\\
&\quad =I_1(u)+I_2(u)+3\cdot (L-u-1),
\end{split}
\end{align}
with
\begin{align*}
I_1(u)&=\int^{L-u}_1\left (\frac{z+u}{z}\right )^{\frac{2}{3}}+2\cdot
\left (\frac{z+u}{z}\right )^{\frac{1}{3}}dz,\\
I_2(u)&=\int^{L-u}_1\left (\frac{z}{z+u}\right )^{\frac{2}{3}}+2\cdot
\left (\frac{z}{z+u}\right )^{\frac{1}{3}}dz.
\end{align*}
To calculate $I_1(u)$ we introduce the change of variables $t=(\tfrac
{z+u}{z})^{\tfrac{1}{3}}$.  Then
\pagebreak
\begin{align}
\begin{split}
&I_1(u)=\int^{(1+u)^{\tfrac{1}{3}}}_{(\tfrac{L}{L-u})^{\tfrac{1}{3}}}
(t^2+2t)\cdot\left (\frac{u}{1-t^3}\right )'dt=\left (
(1+u)^{\tfrac{2}{3}}+2\cdot (1+u)^{\tfrac{1}{3}}\right )\\
&\qquad \cdot (-1)-\left (\left (\frac{L}{L-u}\right 
)^{\tfrac{2}{3}}+2\cdot\left (\frac{L}{L-u}\right )^{\tfrac{1}{3}}\right )
\cdot (-L+u)\\
&\qquad +u\int^{(1+u)^{\tfrac{1}{3}}}_{(\tfrac{L}{L-u})^{\tfrac{1}{3}}}
(2t+2)\cdot\frac{1}{t^3-1}dt.
\end{split}
\end{align}
We have 
\begin{align}
\begin{split}
&u\cdot\int^{(1+u)^{\tfrac{1}{3}}}_{(\tfrac{l}{L-u})^{\tfrac
{1}{3}}}(2t+2)\cdot\frac{1}{t^3-1}dt=u\cdot
\int^{(1+u)^{\tfrac{1}{3}}}_{(\tfrac{L}{L-u})^{\tfrac{1}{3}}}
\frac{4}{3}\cdot\left (\frac{1}{t-1}-\frac{t+\tfrac{1}{2}}{t^2+t+1}
\right )dt\\
&\qquad =\frac{4u}{3}\cdot\log \left [(1+u)^{\tfrac{1}{3}}-1\right ]
-\frac{4u}{3}\log\left [\left (\frac{L}{L-u}\right )^{\tfrac{1}{3}}-1
\right ]\\
&\qquad -\frac{2u}{3}\log\left [(1+u)^{\tfrac{2}{3}}+(1+u)^{\tfrac{1}{3}}
+1\right ]\\
&\qquad +\frac{2u}{3}\log\left [\left (\frac{L}{L-u}\right )^{\tfrac{2}
{3}}+\left (\frac{L}{L-u}\right )^{\tfrac{1}{3}}+1\right ] \ \ \rm{ and}
\end{split}
\end{align}
\begin{align}
\begin{split}
&I_1(u)=-(1+u)^{\tfrac{2}{3}}-2(1+u)^{\tfrac{1}{3}}+L\cdot\left (
1-\frac{u}{L}\right )^{\tfrac{1}{3}}+2L\cdot\left (1-\frac{u}{L}\right 
)^{\tfrac{2}{3}}\\
&\qquad +\frac{4u}{3}\log\left [(1+u)^{\tfrac{1}{3}}-1\right ]-\frac{4u}
{3}\log\left [\left (\frac{L}{L-u}\right )^{\tfrac{1}{3}}-1\right ]\\
&\qquad +\frac{2u}{3}\log\left [\left (\frac{L}{L-u}\right )^{\tfrac{2}{3}}
+\left (\frac{L}{L-u}\right )^{\tfrac{1}{3}}+1\right ]\\
&\qquad -\frac{2u}{3}\log \left [(1+u)^{\tfrac{2}{3}}+(1+u)^{\tfrac{1}{3}}+1
\right ].
\end{split}
\end{align}
\pagebreak
In a similar way in order to calculate
$$I_2(u)=\int^L_{1+u}\left (\frac{z-u}{z}\right )^{\tfrac{2}{3}}
+2\cdot\left (\frac{z-u}{z}\right )^{\tfrac{1}{3}}dz,$$
we consider the change of variables $t=(\tfrac{z-u}{z})^{\tfrac{1}{3}}$.
Then
\allowdisplaybreaks{
\begin{align}
\begin{split}
&I_2(u)=\int^{\left (\frac{L}{L-u}\right
)^{-\tfrac{1}{3}}}_{(1+u)^{-\tfrac{1}{3}}}
(t^2+2t)\cdot\left (\frac{u}{1-t^3}\right )'dt=
\biggl (\left (  \frac{L-u}{L}\right )^{\tfrac{2}{3}}+\\
&\qquad 2\cdot
\left (\frac{L-u}{L}\right )^{\tfrac{1}{3}}\biggr )\cdot L
-\left ((1+u)^{-\tfrac{2}{3}}+2\cdot (1+u)^{-\tfrac{1}{3}}\right )
\cdot (1+u) +\\
&\qquad u\cdot\int^{\left (\tfrac{L}{L-u}\right )^{-\tfrac{1}{3}}}
_{(1+u)^{-\tfrac{1}{3}}}(2t+2)\cdot\frac{1}{t^3-1}dt=-(1+u)^{\tfrac{1}{3}}
-2(1+u)^{\tfrac{2}{3}}\\
&\qquad +L\cdot\left (1-\frac{u}{L}\right )^{\tfrac{2}{3}}+2L\cdot
\left (1-\frac{u}{L}\right )^{\tfrac{1}{3}}+\frac{4}{3}u\cdot\log
\biggl [1-\left (\frac{L-u}{L}\right )^{\tfrac{1}{3}}\biggr ]\\
\end{split}
\end{align}}
\begin{align*}
\begin{split}
&\qquad -\frac{4}{3}u\cdot\log\left [1-(1+u)^{-\tfrac{1}{3}}\right ]
-\frac{2}{3}u\cdot\log\biggl [\left (\frac{L-u}{L}\right )^{\tfrac{2}{3}}\\
&\qquad +\left (\frac{L-u}{L}\right )^{\tfrac{1}{3}}
+1\biggr ]+\frac{2}{3}u\cdot\log\left [(1+u)^{-\tfrac{2}{3}}
+(1+u)^{-\tfrac{1}{3}}+1
\right ].
\end{split}
\end{align*}
It follows from (3.13) that
\begin{align}
\begin{split}
&\int^{-1}_{-L}\int^{-1}_{-L}\left (Q^{(1)}_{0,0}(z_1,z_2)
\right )^2dz_1dz_2=\\
&\qquad \frac{2}{9}\cdot\int^{L-1}_0\left (
\frac{\sin\pi u}{\pi u}\right )^2\cdot (I_1(u)+I_2(u)+3L-3u-3)du .
\end{split}
\end{align}
We note that
\begin{equation}
\int^{L-1}_0 \left (\frac{\sin\pi u}{\pi u} \right )^2\cdot Ldu=\frac
{L}{2}+\underline{0}(1);
\end{equation}

\begin{equation}
\int^{L-1}_0 \frac{(\sin\pi u)^2}{\pi u}du=\frac
{1}{2\pi^2}\log L+\underline{0}(1);
\end{equation}

\begin{align}
\begin{split}
&\int^{L-1}_0\left (\frac{\sin\pi u}{\pi u}\right )^2\cdot L\cdot
\left (1-\frac{u}{L}\right )^{\tfrac{1}{3}}du=L\cdot\int^{L-1}_0
\left (\frac{\sin\pi u}{\pi u}\right )^2\\
&\qquad \cdot\left (1-\frac{1}{3}\frac{u}{L}+\underline{0}\left (\frac
{u^2}{L^2}\right )\right )du=\frac{L}{2}-\frac{1}{6\pi^2}
\log L+\underline{0}(1);
\end{split}
\end{align}

\begin{equation}
\int^{L-1}_0\left (\frac{\sin\pi u}{\pi u}\right )^2\cdot L\cdot
\left (1-\frac{u}{L}\right )^{\tfrac{2}{3}}du=\frac{L}{2}-\frac{1}{3\pi^2}
\log L+\underline{0}(1).
\end{equation}
The combined contribution of all other terms to (3.18) is $\underline{0}
(1)$.  Indeed,
\begin{align*}
\begin{split}
&\int^{L-1}_0\left( \frac{\sin\pi u}{\pi u}\right )^2\cdot 
(1+u)^{\tfrac{2}{3}}=\underline{0}(1);\\
&\int^{L-1}_0\left( \frac{\sin\pi u}{\pi u}\right )^2\cdot 
(1+u)^{\tfrac{1}{3}}=\underline{0}(1);
\end{split}
\end{align*}
\begin{align*}
\begin{split}
&\int^{L-1}_0\left( \frac{\sin\pi u}{\pi u}\right )^2\cdot 
\frac{4u}{3}\log\left [(1+u)^{\tfrac{1}{3}}-1\right ]du-
\int^{L-1}_0\left( \frac{\sin\pi u}{\pi u}\right )^2\\
&\qquad\cdot \frac{2u}{3}\cdot\log\left [(1+u)^{\tfrac{2}{3}}+
(1+u)^{\tfrac{1}{3}}+1\right ]du= \\ &\qquad \frac{2}{3\pi^2}
\int^{L-1}_0\frac{\sin^2\pi u}{\pi^2 u}
\cdot\log\left [\frac{\left ((1+u)^{\tfrac{1}{3}}-1
\right )^2}{(1+u)^{\tfrac{2}{3}}+(1+u)^{\tfrac{1}{3}}+1}
\right ]du= \\ &\qquad \frac{2u}{3\pi^2}\int^{L-1}_0\frac{\sin^2\pi u}{u}
\cdot\underline{0}\left (u^{-\tfrac{1}{3}}\right )
du=\underline{0}(1);
\end{split}
\end{align*}

\begin{equation*}
\int^{L-1}_0\left( \frac{\sin\pi u}{\pi u}\right )^2\cdot
\frac{4u}{3}\log\left [1-(1+u)^{-\tfrac{1}{3}}\right ]du=\underline{0}(1);
\end{equation*}

\begin{equation*}
\int^{L-1}_0\left( \frac{\sin\pi u}{\pi u}\right )^2\cdot
\frac{2u}{3}\log\left [(1+u)^{-\tfrac{2}{3}}+(1+u)^{-\tfrac{1}{3}}+1
\right ]du=\underline{0}(1).
\end{equation*}
The last expression to consider is

\begin{align*}
\begin{split}
	&-\int^{L-1}_0\left( \frac{\sin\pi u}{\pi u}\right )^2\cdot
\frac{4}{3}u\cdot\log\left [\left (\frac{L}{L-u}\right )^{\tfrac{1}{3}}-1
\right ]du+\int^{L-1}_0\left( \frac{\sin\pi u}{\pi u}\right )^2\\
	&\qquad\cdot\frac{2}{3}u\cdot\log\left [\left (\frac{L}{L-u}
\right )^{\tfrac{2}{3}}+\left (\frac{L}{L-u}\right )^{\tfrac{1}{3}}+1
\right ]du+\int^{L-1}_0\left( \frac{\sin\pi u}{\pi u}\right )^2\cdot
\frac{4}{3}u\\
	&\qquad\cdot\log\left [1-\left (\frac{L-u}{L}\right )^{\tfrac{1}{3}}
\right ]du-\int^{L-1}_0\left( \frac{\sin\pi u}{\pi u}\right )^2\cdot
\frac{2}{3}u\cdot\log\Biggl \lbrack
\left (\frac{L-u}{L}\right )^{\tfrac{2}{3}}\\
	&\qquad +\left (\frac{L-u}{L}\right )^{\tfrac{1}{3}}+1
\Biggr \rbrack du=\frac{2}{3\pi^2}\int^{L-1}_0\frac{\sin^2\pi u}{u}
\cdot\log\left [
\tfrac{\left (1-\left (\frac{L-u}{L}\right )^{\tfrac{1}{3}}
\right )}{\left (\left (\frac{L}{L-u}\right )^{\tfrac{1}{3}}-1\right )}
\right .\\
	&\qquad\left .
\times\frac{\left (\left (\frac{L}{L-u}\right )^{\tfrac{2}{3}}+
\left (\frac{L}{L-u}\right )^{\tfrac{1}{3}}+1\right )}
{\left (\left (\frac{L-u}{L}\right )^{\tfrac{2}{3}}+
\left (\frac{L-u}{L}\right )^{\tfrac{1}{3}}+1\right )}
\right ] du=
\frac{2}{3\pi^2}\int^{L-1}_0\frac{\sin^2\pi u}{u}
\cdot\log\left (\frac{L}{L-u}\right )du\\
	&\qquad =\frac{1}{3\pi^2}\cdot\int^{L-1}_1\frac{1}{u}\cdot
\left (-\log (1-\frac{u}{L})\right )du+\underline{0}(1)
=\sqrt{\frac{1}{3\pi^2}}\cdot\int^{L-1}_1\frac{1}{u}\cdot
\sum^\infty_{k=1}\frac{1}{k}\cdot\\
	&\qquad \cdot\left (\frac{u}{L}\right )^kdu+\underline{0}(1)
=\frac{1}{3\pi^2}\cdot\sum^\infty_{k=1}\frac{1}{k^2}+\underline{0}(1)
=\underline{0}(1).
\end{split}
\end{align*}
Combining all the above integrals and looking specifically for the
contributions from (3.19)--(3.22) we obtain
\begin{align*}
\begin{split}
&\int^{-1}_{-L}\int^{-1}_{-L}\left (Q^{(1)}_{0,0}(z_1,z_2)
\right )^2dz_1dz_2=\frac{2}{9}\biggl (3\cdot\frac{L}{2}-3\cdot\frac
{1}{2\pi^2}\log L+\frac{L}{2}-\frac{1}{6\pi^2}\cdot\\
&\qquad \log L+2\cdot\frac{L}{2}-2\cdot\frac{1}{3\pi^2}\cdot\log L
+\frac{L}{2}-\frac{1}{3\pi^2}\log L+2\cdot\frac{L}{2}\\
&\qquad -2\cdot\frac{1}{6\pi^2}\log L+\underline{0}(1)\biggr )=L
-\frac{2}{3\pi^2}\log L+\underline{0}(1).
\end{split}
\end{align*}
\qed 

\pagebreak

In the next lemma we evaluate integrals involving $Q^{(2)}_{0,0}$.
\medskip

\noindent{\bf Lemma 5}  

a) $\int^{-1}_{-L}\int^{-1}_{-L}\left (Q^{(2)}_{0,0}(z_1,z_2)
\right )^2dz_1dz_2=\frac{1}{18\pi^2}\log L+\underline{0}(1)$.

b) $\int^{-1}_{-L}\int^{-1}_{-L} Q^{(1)}_{0,0}(z_1,z_2)\cdot
Q^{(2)}_{0,0}(z_1,z_2)dz_1dz_2=\underline{0}(1)$.
\medskip

\noindent{\it Proof.}  The integral in part a) can be written as
\begin{align*}
\begin{split}
&\frac{1}{9}\int^L_1\int^L_1\left (\frac{\cos\pi (z_1+z_2)}
{\pi (z_1+z_2)}
\right )^2\cdot\left (\frac{z_1^{\tfrac{2}{3}}-z_1^{\tfrac{1}{3}}
\cdot z_2^{\tfrac{1}{3}}+z_2^{\tfrac{2}{3}}}
{z_1^{\tfrac{1}{3}}\cdot z_2^{\tfrac{1}{3}}}\right ) ^2 dz_1dz_2\\
&\qquad =\frac{1}{9}\int^{2L}_2\left (\frac{\cos\pi u}{\pi u}\right )^2
\cdot\int^u_1\biggl [\left (\frac{z}{u-z}\right )^{\tfrac{1}{3}}-1
+\left (\frac{u-z}{z}\right )^{\tfrac{1}{3}}\biggr ]^2dzdu.
\end{split}
\end{align*}
We denote the inner integral by
\begin{align*}
\begin{split}
I_3(u):&=\int^u_1\left [\left (\frac{z}{u-z}\right )^{\tfrac{2}{3}}\!\! -2
\left (\frac{z}{u-z}\right )^{\tfrac{1}{3}}\!\! +\left (\frac{u-z}{z}\right 
)^{\tfrac{2}{3}}\!\! -2\left (\frac{u-z}{z}\right )^{\tfrac{1}{3}}\!\! +1
\right ]dz\\
& =2\cdot\int^u_1\left (\frac{u-z}{z}\right )^{\tfrac{2}{3}}-2\left (
\frac{u-z}{z}\right )^{\tfrac{1}{3}}dz+(u-1)\\
& =2\cdot\int^0_{(u-1)^{\tfrac{1}{3}}}(t^2-2t)\cdot
\left (\frac{u}{t^3+1}\right )'dt+(u-1)\\
&=6u\cdot\int^{(u-1)^{\tfrac{1}{3}}}_0\frac{(t^2-2t)\cdot t^2}
{(t^3+1)^2}dt+(u-1)\\
&=u\left (1+6\int^\infty_0\frac{t^4-2t^3}
{(t^3+1)^2}dt\right )+\underline{0}(u^{\tfrac{2}{3}}).
\end{split}
\end{align*}
Taking into account
$\int^\infty_0\frac{t^4-2t^3}{(t^3+1)^2}dt=0$, we obtain
$I_3(u)=u+\underline{0}(u^{\tfrac{2}{3}})$, and 
$$\frac{1}{9}
\int^{2L}_2(\frac{\cos\pi u}{\pi u})^2\cdot (u+\underline{0}
(u^{\tfrac{2}{3}}))du=\frac{1}{18\pi^2}\log L+\underline{0}(1).$$

Now let us consider the integral in part b):
\begin{align*}
\begin{split}
	&\frac{1}{9}\int^L_1\int^L_1\frac{\cos\pi (z_1+z_2)}
{\pi(z_1+z_2)}\cdot\frac{\sin\pi (z_1-z_2)}{\pi\cdot (z_1-z_2)}\\
	&\qquad\biggl (\frac{z_1^{\tfrac{2}{3}}+z_1^{\tfrac{1}{3}}
\cdot z_2^{\tfrac{1}{3}}+z_2^{\tfrac{2}{3}}}
{z_1^{\tfrac{1}{3}}\cdot z_2^{\tfrac{1}{3}}}\biggr )\cdot
\left (\frac{z_1^{\tfrac{2}{3}}-z_1^{\tfrac{1}{3}}
\cdot z_2^{\tfrac{1}{3}}+z_2^{\tfrac{2}{3}}}
{z_1^{\tfrac{1}{3}}\cdot z_2^{\tfrac{1}{3}}}\right )dz_1dz_2=\\
	&\qquad =\frac{1}{9}\cdot\int^{2L}_2\frac{\cos\pi u}{\pi u}
\cdot\int^{u-2}_0\frac{\sin\pi v}{\pi v}\cdot\biggl [
\left (\frac{u+v}{u-v}\right )^{\tfrac{2}{3}}+\left (\frac{u-v}
{u+v}\right )^{\tfrac{2}{3}}+1\biggr ]dvdu.
\end{split}
\end{align*}
It is not difficult to see that oscillations of trigonometric functions
make this integral to be of order of constant.  To show this we write the
inner integral as $I_4(u)+I_5(u)+I_6(u)$, where
\begin{align*}
I_4(u)&=\int^{u-2}_0\frac{\sin\pi v}{\pi v}\cdot\left (\frac{u+v}
{u-v}\right )^{\tfrac{2}{3}}dv,\\
I_5(u)&=\int^{u-2}_0\frac{\sin\pi v}{\pi v}\cdot\left (\frac{u+v}
{u-v}\right )^{\tfrac{1}{3}}dv,\\
I_6(u)&=\int^{u-2}_0\frac{\sin\pi v}{\pi v}dv.
\end{align*}
Integration by parts gives $I_6(u)=\frac{1}{2}+\underline{0}(u^{-1})$.
Consider now 
\begin{equation}
I_4(u)=\int^{u-2u^{\tfrac{1}{4}}}_0\left (\frac{\sin\pi v}{\pi v}
\right )\cdot\left (\frac{u+v}{u-v}\right )^{\tfrac{2}{3}}dv+\int
\limits^{u-2}_{u-2u^{\tfrac{1}{4}}}\left (\frac{\sin\pi v}{\pi v}\right )
\cdot\left (\frac{u+v}{u-v}\right )^{\tfrac{2}{3}}dv.
\end{equation}
The second integral in (3.23) is $\underline{0}(u^{\tfrac{1}{4}}
\cdot u^{-1}\cdot u^{\tfrac{2}{3}})=\underline{0}(u^{-\tfrac{1}{12}})$.
As for the first one we introduce $\chi ([v]$ even), an indicator of the
set where the integer part of $v$ is even, and write it in the following
form
\begin{align*}
\begin{split}
	&\int^{u-2u^{\tfrac{1}{4}}}_0\frac{\sin\pi v}{\pi}\cdot\chi ([v]\ 
\text{even})\cdot\biggl (\frac{1}{v}\cdot\left (\frac{u+v}{u-v}
\right )^{\tfrac{2}{3}}-\frac{1}{v+1}\\
	&\qquad \left (\frac{u+v+1}{u-v-1}\right )^{\tfrac{2}{3}}\biggr )dv
+\underline{0}(u^{-\tfrac{1}{2}}).
\end{split}
\end{align*}
The absolute value of the last expression is estimated from above by
\begin{align*}
\begin{split}
	&\int^{u-2u^{\tfrac{1}{4}}}_0\frac{\vert\sin\pi v\vert}
{\pi}\cdot\left (\frac{v\cdot (u-v)^{\tfrac{2}{3}}-(u-v-1)^{\tfrac{2}{3}}
\cdot (v+1)}
{v\cdot (v+1)\cdot (u-v)^{\tfrac{2}{3}}\cdot (u-v-1)^{\tfrac{2}{3}}}
\right )dv\\
	&\qquad =\int^{u-2u^{\tfrac{1}{4}}}_0\frac{\vert\sin\pi 
v\vert}{\pi}\cdot\left (\frac{(u-v)^{\tfrac{2}{3}}-(u-v-1)^{\tfrac{2}{3}}}
{(v+1)\cdot (u-v)^{\tfrac{2}{3}}\cdot (u-v-1)^{\tfrac{2}{3}}}\right )dv\\
	&\qquad +\int^{u-2u^{\tfrac{1}{4}}}_0\frac{\vert\sin\pi 
v\vert}{\pi}\cdot\left (\frac{1}{v\cdot (v+1)\cdot (u-v)^{\tfrac{2}{3}}}
\right )dv=\underline{0}(u^{-\tfrac{1}{6}}).
\end{split}
\end{align*}
Therefore $I_4(u)=\underline{0}(u^{-\tfrac{1}{6}})$, and similarly
$I_5(u)=\underline{0}(u^{-\tfrac{1}{12}})$.  As a result
\begin{align*}
\begin{split}
&\frac{1}{9}\int^{2L}_2\frac{\sin\pi u}{\pi u}(I_4(u)+I_5(u)+
I_6(u))du =\frac{1}{18}\int^{2L}_2\frac{\sin\pi u}{\pi u}du\\
&\qquad +\frac{1}{9}\int^{2L}_2\frac{\sin\pi u}{\pi u}\cdot
\underline{0}(u^{-\tfrac{1}{12}})du=\underline{0}(1).
\end{split}
\end{align*}
Lemma 5 is proven.\qed

As a result of the last two lemmas we have
\begin{align*}
\begin{split}
&\int^{-1}_{-L}\int^{-1}_{-L}
S^2(z_1,z_2)dz_1dz_2=L-\frac{2}{3\pi^2}\log L+\frac{1}
{18\pi^2}\log L+\underline{0}(1)=\\
&\qquad L-\frac{11}{18\pi^2}\log L+\underline{0}
(1).
\end{split}
\end{align*}
Remember that $S$ was defined as $S(z_1,z_2)=Q^{(1)}_{0,0}(z_1,z_2)
+Q^{(2)}_{0,0}(z_1,z_2)$.  To finish the proof of the CLT 
for $ \# (-T, +\infty ) $ we just need
to show that the remainder term $U(z_1,z_2)=Q(z_1,z_2)-S(z_1,z_2)$
is insignificant in the following sense:
\medskip

\noindent{\bf Lemma 6}

a) $\int^{-1}_{-L}\int^{-1}_{-L}U^2(z_1,z_2)dz_1dz_2=
\underline{0}(1)$

b) $\int^{-1}_{-L}\int^{-1}_{-L} U(z_1,z_2)\cdot S(z_1,z_2)
dz_1dz_2=\underline{0}(1)$

c) $\int^{-1}_{-L} U(z,z)dz=\underline{0}(1)$

\medskip

\noindent{\it Proof.}  We shall establish part a).  
Parts b) and c) can be treated in a similar
manner.  Repeating the calculations of the last two lemmas, it is easy to see
that for any fixed indices $(i,m,n)$ such that $(i-1)(i-2)+m+n>0$, we have
$$\int^{-1}_{-L}\int^{-1}_{-L}\left (Q^{(i)}_{m,n}
(z_1,z_2)\right )^2dz_1dz_2=\underline{0}(1).$$
Let us now choose $N$ to be sufficiently large and write $U(z_1,z_2)=
U_N(z_1,z_2)+V_N(z_1,z_2)$, where
$$U_N(z_1,z_2)=\sum^2_{i=1}\sum_{0<m+n\leq N}Q^{(i)}_{m,n}(z_1,z_2)
+\sum^6_{i=3}\sum_{0\leq m+n\leq N}Q^{(i)}_{m,n}(z_1,z_2).$$
We see that $\int^{-1}_{-L}\int^{-1}_{-L}
(U_N(z_1,z_2))^2dz_1dz_2=\underline{0}(1)$.  Asymptotic formulas (3.6)--(3.11)
imply that
\begin{align}
\begin{split}
&\vert V_N(z_1,z_2)\vert\leq\frac{\const_N}{\vert z_1^{\tfrac{2}{3}}
-z_2^{\tfrac{2}{3}}\vert}\cdot\left (
z_1^{-2N}+z_n^{-2N}\right )\\
&\qquad \leq\frac{\const_N}{\vert z_1-z_2\vert}
\cdot\frac{3}{2}\cdot\left (z^{\tfrac{1}{3}}_1+z^{\tfrac{1}{3}}_2\right )
\cdot\left ( z_1^{-2N}+z_2^{-2N}\right ).
\end{split}
\end{align}
It follows from (3.24) that if we choose $N\geq 2$ then
\begin{equation}
\iint\limits_{\vert z_1-z_2\vert\geq\tfrac{1}{z^2_2}}\left (V_n(z_1,z_2)
\right )^2dz_1dz_2=\underline{0}(1).
\end{equation}
(The integration in (3.25) is over the subset of $[1,L]\times [1,L]$) . 
Indeed, to estimate the integral over $z_2$ we write
\begin{align*}
\begin{split}
\int_{\vert z_1-z_2\vert \geq\tfrac{1}{z^2_2}}\left (\frac{z_1}
{(z_1-z_2)}\cdot\frac{1}{z_2^{2N}}\right )^2dz_2 &\leq
\int^L_1\frac{z^4_2}{(z_1-z_2)^2+1}\cdot z_1^{\tfrac{2}{3}}\cdot
\frac{1}{z_2^{4N}}dz_2\\
&\leq \int^L_1\frac{z_1^{\tfrac{2}{3}}}{(z_1-z_2)^2+1}
\cdot z^{-4}_2dz_2=\underline{0}(z_1^{-\tfrac{4}{3}}).
\end{split}
\end{align*}
Integrating over $z_1$ we arrive at (3.25).
To integrate $V^2_N$ near the diagonal we observe that kernels 
$Q(z_1,z_2)$, $Q^{(i)}_{m,n}(z_1,z_2)$ are bounded in $[1,+\infty )\times
[1,+\infty )$; therefore there exists some $\const_N'$ such that
$\vert V_N(z_1,z_2)\vert\leq\const_N'$ and $\iint_{\vert z_1-z_2
\vert\leq\tfrac{1}{z^2_2}}(V_N(z_1,z_2))^2dz_1dz_2\leq\int^{+\infty}_1
\tfrac{2\cdot\const_N'}{z^2_2}dz_2=\underline{0}(1)$.
Lemma 6 is proven.\qed
\medskip

Taking $L=\tfrac{2}{3\pi}T^{\tfrac{3}{2}}$ we deduce from Lemmas 4-6 and
(3.5) that
\begin{align*}
\begin{split}
\var \biggl (\#\biggl (y_i\in \biggl (-T,-\biggl (\frac{3\pi}{2}
\biggr )^{\tfrac{2}{3}}\biggr )
\biggr )&=\int^{-\left (\tfrac{3\pi}{2}\right )^{\tfrac{2}{3}}}_{-T}
K(y,y)dy\\
&\quad -\int^{-\left (\tfrac{3\pi}{2}\right )^{\tfrac{2}{3}}}_{-T}
\int^{-\left (\tfrac{3\pi}{2}\right )^{\tfrac{2}{3}}}_{-T}
K^2(y_1,y_2)dy_1dy_2\\
&=\frac{2}{3\pi}T^{\tfrac{2}{3}}+\underline{0}(1)
-\frac{2}{3\pi}T^{\tfrac{2}{3}}\\
&\quad +\frac{11}{18\pi^2}\log \left (\frac{2}{3\pi}T^{\tfrac{3}{2}}\right )
+\underline{0}(1)\\
&=\frac{11}{12\pi^2}\log T+\underline{0}(1).
\end{split}
\end{align*}
This finishes the proof of Proposition 2 as well as the proof of the CLT for 
$ \# (-T, +\infty ) $.\\
In a very similar way one proves the CLT for arbitrary $\ \nu_k(T) =
\# \left ( ( -kT, -(k-1)T) \right ),\\  \ k>1. \ $ To prove the result for the
joint distribution of $ \{ \nu_k(T) \ \} $ we note that the decay of $ \ K(x,y)
\ $ off the diagonal implies
$$ C_{1,1}(\nu_k(T), \nu_l(T)) = \rm Cov \left (\nu_k(T), \nu_l(T) \right  )
= \int_{-kT}^{-(k-1)T} \int_{-lT}^{-(l-1)T} \ K^2(x_1, x_2) dx_1, dx_2 = $$
$$-\Trace \chi_{[-lT, -(l-1)T)} \cdot K \cdot \chi_{[-kT, -(k-1)T)}\cdot K =
\underline{O}(1) \ \ 
\rm{if}  \ \  |k-l|>1 .$$ This together with
$$ \var \left ( \sum_{l=1}^k \nu_l(T) \right )
=11/(12 \pi^2) \log T +
\underline{O}(1), \ \ \ \var (\nu_k(T))=11/(12 \pi^2) \log T +
\underline{O}(1), \ \ , k=1,2,\ldots $$
implies that
$$ C_{1,1}(\nu_k(T), \nu_l(T)) = \rm Cov \left (\nu_k(T), \nu_l(T) \right  )=$$
$$11/(12 \pi^2) \log T +
\underline{O}(1)  $$ for $\ \ |k-l|=1.\ \ $  Therefore as 
$ \ \ T \to \infty \ \ $.
$$ E \frac{\nu_k (T)-E\nu_k (T)}{\sqrt {\var \ \nu_k (T)}}\cdot
  \frac{\nu_l (T)-
E\nu_l (T)}{\sqrt {Var\ \nu_l (T)}} \to \delta_{k,l}- 1/2 \ \delta_{k,l-1}-
1/2 \ \delta_{k,l+1} $$
To take care of the joint cumulants of higher order it is enought to prove
\medskip

\noindent{\bf Lemma 7}  {\it
Let at least two indices in $(k_1,\ldots, \k_s)$ are non-zero. Then
$$ C_{k_1,\ldots, k_s} \left ( \nu_1(T),\ldots , \nu_s(T)\right ) =
\underline{O}(\log T)$$
.}
\medskip

\noindent{\it  Proof}
According to Proposition 1 Lemma 7 follows from
\medskip

\noindent{\bf Lemma 8}  {\it
$$ \Trace \chi_{l_1} K \chi_{l_2} K \ldots K \chi_{l_s} K \chi_{l_1} =
\underline{O}(\log T)
$$
where $\chi_{l_j}$ are  the indicators of the intervals $ (-l_j T, -(l_j-1)T]
, \ l_j \in Z^1_{+} $  and at least two intervals are disjoint
.}
\medskip

\noindent{ \it Proof}
This has been already established for $s=2$. Let $ \ s >2 \ .$ Since not all indices coincide by cyclicity of the trace we may assume $\ l_1 \ne l_2 \ .$ Now
if $\ l_1 =l_3 \ $ we can use the positivity of
$\ \ \chi_{l_1} K \chi_{l_2} K \chi_{l_1} \ \ $ to write
$$ \left | \Trace ( \chi_{l_1} K\chi_{l_2} K\chi_{l_1} K\chi_{l_4}\ldots 
 K\chi_{l_s} K\chi_{l_1} ) \right | \leq \Trace \left ( \chi_{l_1} K\chi_{l_2}
 K\chi_{l_1} \right ) \cdot \parallel  K\chi_{l_4}\ldots  K\chi_{l_s} K
\chi_{l_1} \parallel $$
$$ \leq \Trace \left ( \chi_{l_1} K\chi_{l_2} K\chi_{l_1} \right ) $$
where we used $ \parallel K \parallel \leq 1, \ \ \parallel \chi_{l_j}\parallel
\leq 1 .$ Since
$ \ \Trace \left ( \chi_{l_1} K\chi_{l_2} K\chi_{l_1} \right ) = 
\underline{O} (\log T) \ \ $
 Lemma 8 is proven  when $ \ \ l_1 =l_3 \ne l_2 $.\\
If $ \ l_1 \ne l_3 \ $ one more trick is needed. Let  us denote 
$$ D_1 = \chi_{l_1} K \chi_{l_2}, \ \ 
D_2 =K\chi_{l_3} K \ldots \chi_{l_s} K \chi_{l_1} . $$
Then
$$ \Trace \left ( \chi_{l_1} K\chi_{l_2} K\chi_{l_3}
 \ldots \chi_{l_s}K\chi_{l_1}
\right )
=\Trace (D_1 D_2)$$ 
$$ \leq \left (\Trace (B_1 B_1^*)\right )^{1/2}
\left (\Trace (D_2 D_2^*)\right )^{1/2} $$
(see  [RS], volume I, section VI.6). As before
$$ \Trace (D_1 D_1^*) =  \Trace \left ( \chi_{l_1} K\chi_{l_2} K\chi_{l_1}
\right ) =\underline{O}(\log T) $$
To obtain a similar bound for $\ \Trace (D_1 D_1^*)  $  we define $\ 1<p\leq s
\ $ as the maximal index such that $\ \ l_p \ne l_1 \ .$ Since we assume in 
Lemma 8 that there are at least two different indices, such $p $ always exists.
Then
$$ \Trace \left (K \chi_{l_3} K\ldots \chi_{l_p} K\chi_{l_1}\ldots K\chi_{l_1}
\ldots K\chi_{l_1} \right )\cdot \left (K \chi_{l_3} K\ldots \chi_{l_p} K
\chi_{l_1}\ldots K\chi_{l_1}\ldots K\chi_{l_1}\right )^{*} =$$
$$  \Trace \left (K \chi_{l_3} K\ldots \chi_{l_p} K\chi_{l_1}\ldots K\chi_{l_1}
\ldots K\chi_{l_1} \right )\cdot \left (\chi_{l_1} K\ldots \chi_{l_1} K\ldots K\chi_{l_1}K \chi_{l_p} \ldots K \chi_{l_3} K \right ). $$
Using the identity $ \ \ \Trace (D_1 D_2)= \Trace (D_2 D_1) \ \ $ where
$$ D_1= K\chi_{l_3} K\ldots \chi_{l_{p-1}} K, \ \ 
 D_2= \chi_{l_p} K \chi_{l_1} K \ldots \chi_{l_1} K \chi_{l_1} 
 \ldots \chi_{l_1} K\chi_{l_p} K \chi_{l_{p-1}} K \ldots K
\chi_{l_3} K , $$
we can rewrite and estimate the r.h.s. as
$$ \left | \Trace \left (\chi_{l_p} K \chi_{l_1}\ldots \chi_{l_1}K
\ldots  \chi_{l_1}K \chi_{l_p}
\right )\cdot \left (K \chi_{l_{p-1}}\ldots K\chi_{l_3} K K\chi_{l_3} K \ldots
\chi_{l_{p-1}} K \right ) \right |$$
$$\leq \Trace \left (\chi_{l_p}K\chi_{l_1}\ldots K\chi_{l_1}K\ldots 
\chi_{l_1}K\chi_{l_p}\right )  \cdot \parallel K\chi_{l_{p-1}}\ldots K
\chi_{l_3}K K \chi_{l_3}K\ldots \chi_{l_{p-1}} K \parallel $$  
Here we used the positivity of $\ \ \ \chi_{l_p}K\chi_{l_1}K\chi_{l_1}\ldots
K\chi_{l_1}\ldots K\chi_{l_1}\ \ \ .$  
The norm of the last factor is again not greater than $1$.
Finally,
$$  \Trace \left ( \chi_{l_p} K \chi_{l_1} K \chi_{l_1} \ldots K\chi_{l_1}
\ldots \chi_{l_1}K \chi_{l_p}\right )=\Trace \left ( \chi_{l_1} K \chi_{l_p}
 K \chi_{l_1} \ldots K\chi_{l_1}\ldots K\chi_{l_1}\right ) \leq$$
$$ \Trace \left (\chi_{l_1}K\chi_{l_p}K\chi_{l_1} \right ) \cdot 1 =
\underline{O}(\log T). $$
Here we also used cyclicity of the trace.
Combining the estimates for $ \ \Trace D_1 D_1^* \ \ $ and $ \ \Trace D_1 
D_1^* \ \ $ we finish the proof of the Lemmas $7$ and $ 8.$ \qed \\
It follows from Lemma 7 that the higher joint cumulants of the normalized 
random variables go to zero which implies that the limiting distribution 
function is gaussian with the known covariance function. 
Theorem 1 is proven . \qed

\section{Proof of Theorem 2 and Similar\\Results for the Classical Compact 
Groups}

The Bessel kernel has the form (see \S 1)
\begin{align}
\begin{split}
&K(y_1,y_2)=\frac{J_\alpha (\sqrt{y_1})\cdot \sqrt{y_2}\cdot J'_\alpha 
(\sqrt{y_2})-\sqrt{y_1}\cdot J'_\alpha (\sqrt{y_1})\cdot J_\alpha
(\sqrt{y_2})}{2(y_1-y_2)},\\
&\qquad y_1,y_2\in (0,+\infty ),\ \alpha >-1.
\end{split}
\end{align}
The level density is given by
\begin{equation}
\rho_1(y)=K(y,y)=\frac{1}{4} J_\alpha (\sqrt{y})^2-\frac{1}{4} J_{\alpha +1}
(\sqrt{y})\cdot J_{\alpha -1}(\sqrt{y}).
\end{equation}
The asymptotic formula for large $y$ is well known in 
the case of Bessel functions
(see asymptotic expansion in (4.5) below).  In particular, one can see
that
\begin{equation}
\rho_1(y)\sim\frac{1}{2\pi \sqrt{y_1}}\text{ for }y\rightarrow +\infty .
\end{equation}
The last formula suggests to make the (unfolding) change of variables
$z_i=\frac{\sqrt{y_i}}{\pi},\ i=1,2$.  The kernel $Q$ corresponding
to the new evenly spaced random point field is given by
\begin{align}
\begin{split}
Q(z_1,z_2)&=2\pi y_1^{\tfrac{1}{4}}\cdot y_2^{\tfrac{1}{4}}\cdot K(y_1,y_2)\\
&=z_1^{\tfrac{1}{2}}z_2^{\tfrac{1}{2}}\cdot\frac{J_\alpha (\pi z_1)
\cdot \pi z_2\cdot J'_\alpha (\pi z_2)-\pi z_1\cdot J'_\alpha (\pi z_1)
\cdot J_\alpha (\pi z_2)}{z_1^2-z_2^2}.
\end{split}
\end{align}
The asymptotic expansion of the Bessel function at infinity is due to 
Hankel (see, for example, [Ol]).
\begin{align}
\begin{split}
&J_\alpha (z)\sim\left (\frac{2}{\pi z}\right )^{\tfrac{1}{2}}\cdot\biggl [
\cos\left (z-\frac{1}{2}\alpha\pi -\frac{1}{4}\pi \right )\cdot
\sum^\infty_{s=0}(-1)^s\cdot\frac{A_{2s}(\alpha )}{z^{2s}}\\
&\qquad -\sin \left (z-\frac{1}{2}\alpha\pi -\frac{1}{4}\pi\right )
\cdot\sum^\infty_{s=0}(-1)^s\cdot\frac{A_{2s+1}(\alpha )}{z^{2s+1}}\biggr ],
\end{split}
\end{align}
where
\begin{equation}
A_0(\alpha )=1,\ A_s(\alpha )=\frac{(4\alpha^2-1^2)\cdot (4\alpha^2-3^2)
\dots (4\alpha^2-(2s-1)^2)}
{s!\cdot 8^s}.
\end{equation}
Similarly,
\begin{align}
\begin{split}
&J'_\alpha (z)\sim\left (\frac{2}{\pi z}\right )^{\tfrac{1}{2}}\cdot\biggl [
-\sin \left (z-\frac{1}{2}\alpha\pi -\frac{1}{4}\pi \right )\cdot
\sum^\infty_{s=0}(-1)^s\cdot\frac{B_{2s}(\alpha )}{z^{2s}}\\
&\qquad -\cos\left (z-\frac{1}{2}\alpha\pi -\frac{1}{4}\pi\right )\cdot
\sum^\infty_{s=0}(-1)^s\cdot\frac{B_{2s+1}(\alpha )}{z^{2s+1}}
\biggr ].
\end{split}
\end{align}
The coefficients $B_s(\alpha )$ can be obtained from (4.5)--(4.6); for example,
$B_0(\alpha )=1$.

It follows from (4.5)--(4.7) that for $\alpha =\pm\tfrac{1}{2}$
\begin{equation}
Q(z_1,z_2)=\frac{\sin\pi (z_1-z_2)}{\pi (z_1-z_2)}+\frac{\sin\pi (z_1+z_2
-\alpha -\tfrac{1}{2})}{\pi (z_1+z_2)},
\end{equation}
which can be further simplified as
$$\frac{\sin\pi (z_1-z_2)}{\pi (z_1-z_2)}\mp\frac{\sin\pi (z_1+z_2)}
{\pi (z_1+z_2)}.$$
For general values of $\alpha$ a small remainder term appears at the
r.h.s. of (4.8).  To write the asymptotic expansion, we represent $Q(z_1,z_2)$
as the sum of six kernels:  $Q(z_1,z_2)=\sum\limits^6_{i=1} Q^{(i)}
(z_1, z_2)$, where
\begin{align}
\begin{split}
&Q^{(1)}(z_1,z_2)\sim\frac{\sin\pi (z_1-z_2)}{\pi (z_1^2-z^2_2)}\cdot
\sum^\infty_{n,m=0}(-1)^{n+m}\cdot A_{2n}(\alpha )B_{2m}(\alpha )\\n
&\qquad \cdot \left (z_1^{1-2n}\cdot z_2^{-2m}+z_1^{-2m}\cdot z_2^{1-2n}
\right ),
\end{split}
\end{align}

\begin{align}
\begin{split}
&Q^{(2)}(z_1,z_2)\sim\frac{\sin\pi (z_1+z_2-\alpha -\tfrac{1}{2})}
{\pi (z_1^2-z^2_2)}\cdot\sum^\infty_{n,m=0}(-1)^{n+m}\cdot A_{2n}(\alpha )\\
&\qquad \cdot B_{2m}(\alpha )\cdot \left (z_1^{1-2n}\cdot z_2^{-2m}-
z_1^{-2m}\cdot z_2^{1-2n}\right ),
\end{split}
\end{align}

\begin{align}
\begin{split}
&Q^{(3)}(z_1,z_2)\sim\frac{2\cos\left (\pi z_1-\tfrac{\pi\alpha}{2}
-\tfrac{\pi}{4}\right )\cdot\cos\left (\pi z_2-\tfrac{\pi\alpha}{2}
-\tfrac{\pi}{4}\right )}
{\pi (z_1^2-z^2_2)}\\
&\qquad \cdot\sum^\infty_{n,m=0}(-1)^{n+m+1}
\cdot A_{2n}(\alpha )\cdot B_{2m+1}(\alpha )\cdot \left (z_1^{-2n}
\cdot z_2^{-2m}+z_1^{-2m}\cdot z_2^{-2n}\right ),
\end{split}
\end{align}

\begin{align}
\begin{split}
&Q^{(4)}(z_1,z_2)\sim\frac{2\sin\left (\pi z_1-\tfrac{\pi\alpha}{2}
-\tfrac{\pi}{4}\right )\cdot\sin\left (\pi z_2-\tfrac{\pi\alpha}{2}
-\tfrac{\pi}{4}\right )}
{\pi (z_1^2-z^2_2)}\\
&\qquad \cdot\sum^\infty_{n,m=0}(-1)^{n+m}
\cdot A_{2n+1}(\alpha )\cdot B_{2m}(\alpha )\cdot \left (z_1^{-1-2n}
\cdot z_2^{1-2m}+z_1^{1-2m}\cdot z_2^{-1-2n}\right ),
\end{split}
\end{align}

\begin{align}
\begin{split}
&Q^{(5)}(z_1,z_2)\sim\frac{\sin\pi\left (z_1-z_2\right )}
{\pi (z_1^2-z^2_2)}\cdot\sum^\infty_{n,m=0}(-1)^{n+m+1}
\cdot A_{2n+1}(\alpha )\cdot B_{2m+1}(\alpha )\\
&\qquad \cdot \left (z_1^{-1-2n}
\cdot z_2^{-2m}+z_1^{-2m}\cdot z_2^{-1-2n}\right ),
\end{split}
\end{align}

\begin{align}
\begin{split}
&Q^{(6)}(z_1,z_2)\sim\frac{\sin\pi\left (z_1+z_2-\alpha -\tfrac{1}{2}\right )}
{\pi (z_1^2-z^2_2)}\cdot\sum^\infty_{n,m=0}(-1)^{n+m+1}
\cdot A_{2n+1}(\alpha )\\
&\qquad \cdot B_{2m+1}(\alpha )\cdot \left (z_1^{-1-2n}
\cdot z_2^{-2m}-z_1^{-2m}\cdot z_2^{-1-2n}\right ).
\end{split}
\end{align}
The analysis of (4.9)--(4.14) is very similar to \S 3.  One can see that
the only contribution to the leading term of the variance comes from
$Q^{(1)}_{0,0}(z_1,z_2)+Q^{(2)}_{0,0}(z_1,z_2)$ which is exactly the r.h.s.
of (4.8).  It can be shown by a straightforward calculation that
\begin{align*}
\begin{split}
&\int^L_0\left (Q^{(1)}_{0,0}(z,z)+Q^{(2)}_{0,0}(z,z)\right )dz\sim
L+\underline{0}(1),\\
&\qquad\int^L_0\int^L_0
\left (Q^{(1)}_{0,0}(z_1,z_2)+Q^{(2)}_{0,0}
(z_1,z_2))\right )^2dz_1dz_2
\sim L-\frac{1}{2\pi^2}\log L+\underline{0}(1).
\end{split}
\end{align*}
Taking into account that $L=\tfrac{T^{\tfrac{1}{2}}}{\pi}$ we finish the
proof.\qed
\medskip

The kernels $\tfrac{\sin\pi (x-y)}{\pi (x-y)}\pm\tfrac{\sin\pi (x+y)}
{\pi (x+y)}$ are well known in Random Matrix Theory.  For one, they are the
kernels of restrictions of the sine-kernel integral operator to the subspaces
of even and odd functions and play an important role in spacings distribution
in G.O.E. and G.S.E. ([Me]).  They also appear as the kernels of limiting
correlation functions in orthogonal and symplectic groups near $\lambda =1$
([So1]).  Let us start with the even case.  Consider the normalized
Haar measure on $SO(2n)$.
  The eigenvalues of matrix $M$ can be
arranged in pairs:
\begin{equation}
\exp (i\theta_1),\exp (-i\theta_1),\dots ,\exp (i\theta_n),\exp (-i\theta_n),
0\leq\theta_1,\theta_2,\dots ,\theta_n\leq\pi .
\end{equation}
In the rescaled coordinates near the origin $x_i=(2n-1)\cdot\tfrac{\theta_i}
{2\pi},\ i=1,\dots ,n$, the $k$-point correlation functions are equal
to
\begin{align}
\begin{split}
&R_{n,k}(x_1,\dots ,x_k)=\det \biggl (\frac{\sin\pi (x_i-x_j)}{(2n-1)\cdot
\sin (\pi\cdot (x_i-x_j)/(2n-1))}\\
&\qquad +\frac{\sin\pi (x)i+x_j)}{(2n-1)\cdot\sin (\pi (x_i+x_j)/(2n-1))}
\biggr )_{i,j=1,\dots k}.
\end{split}
\end{align}
In the limit $n\rightarrow\infty$ the kernel in (4.16) becomes
$\tfrac{\sin\pi (x_i-x_j)}{\pi (x_i-x_j)}+\tfrac{\sin\pi (x_i+x_j)}
{\pi (x_i+x_j)}$.  If we consider rescaling near arbitrary $0<\theta<\pi$
the limiting kernel will be just the sine kernel.

Let us now consider the $SO(2n+1)$ case.  The first $2n$ eigenvalues of
$M\in SO(2n+1)$ can be arranged in pairs as in (4.15).  The last one equals
1.  In the rescaled coordinates near $\theta =0$
$$x_i=\frac{n\theta_i}{\pi},\ i=1,\dots , n,$$
the $k$-point correlation functions are given by the formula
\begin{align}
\begin{split}
&R_{n,k}(x_1,\dots ,x_k)=\det\biggl (\frac{\sin\pi (x_i-x_j)}{2n\cdot\sin
(\pi\cdot (x_i-x_j)/2n)}\\
&\qquad -\frac{\sin\pi (x_i+x_j)}{2n\cdot\sin (\pi\cdot (x_i+x_j)/2n)}
\biggr )_{i,j=1\dots k}.
\end{split}
\end{align}
In the limit $n\rightarrow\infty$ the kernel in (4.17) becomes 
$\tfrac{\sin\pi (x_i-x_j)}{\pi\cdot (x_i-x_j)}
-\tfrac{\sin\pi (x_i+x_j)}{\pi\cdot (x_i+x_j)}$.
If we again consider rescaling near $0<\theta <\pi$, the limiting
kernel appears to be the sine kernel.  The case of symplectic group is
very similar.  Let $M\in Sp(n)$.  Rescaled $k$-point correlation functions
are given by
\begin{align}
\begin{split}
&R_{n,k}(x_1,\dots , x_k)=\det\biggl (\frac{\sin\pi (x_i-x_j)}{(2n+1)\cdot
\sin (\pi (x_i-x_j)/(2n+1)}\\
&\qquad -\frac{\sin\pi (x_i+x_j)}{(2n+1)\cdot\sin (\pi (x_i+x_j)/(2n+1))}
\biggr )_{i,j=1,\dots k}.
\end{split}
\end{align}

One can then deduce the following result from the Costin-Lebowitz Theorem.
\medskip

\noindent{\bf Theorem 3}  {\it
Consider the normalized Haar measure on $SO(n)$ or $Sp(n)$.  Let $\theta\in
[0,\pi ),\ \delta_n$ be such that $0<\delta_n<\pi -\theta -\epsilon$ for 
some $\epsilon >0$ and $n\cdot\delta_n\rightarrow +\infty$.  Denote by
$\nu_n$ the number of eigenvalues in $[\theta ,\theta +\delta_n]$.  We have
$E\nu_n=\tfrac{n}{\pi}\cdot\delta_n +\underline{0}(1)$,
$$\var\ \nu_n=
\begin{cases}
\frac{1}{\pi^2}\log (n\cdot\delta_n)+\underline
{0}(1) & \text{if }\theta >0,\\ 
\frac{1}{2\pi^2}\log (n\cdot\delta_n)+
\underline{0}(1) &\text{ if }\theta =0
\end{cases}
$$
and the normalized random variable 
$\tfrac{\nu_n-E\nu_n}
{\sqrt{\var\ \nu_n}}$ converges in distribution to the 
normal law
$N(0,1)$.}
\medskip

\noindent{\it Proof}.  Let $K_n(x,y)$ be the kernel in (4.16), (4.17), or (4.18).  We observe
that $0\leq K_n\cdot\xi_I\leq Id$ as a composition of projection,
Fourier transform, another projection and inverse Fourier transform.
To check the asymptotics of \break Var $\nu_n$ is an excercise which is left
to the reader. \qed \\

The case of several intervals is treated in a similar fasion.

\medskip

\noindent{\bf Theorem 4}  {\it
Let $ \delta _n >0 $ be such that $ \delta _n \to 0, \ n\delta_n \to \infty $
and $\nu_{k,n} = \# \left ( (k-1)\delta_n, k\delta_n] \right ). $
Then  a sequence of normalized random variables
$\tfrac{\nu_{k,n}-E\nu_{k,n}}
{\sqrt{\var\ \nu_{k,n}}}$ converges in distribution to the centalized gaussian
sequence $ \{ \xi_k \}_{k=1}^{\infty} \}$ with the covariance function
$$ E \xi_k \xi_l=
\begin{cases}
\delta_{k,l} - 1/2 \ \delta_{k,l+1} -1/2 \ \delta_ {k,l-1},
 & \text{if } k>0, \ l>0\\ 
\delta_{0,l} -1/ \sqrt{2} \ \delta_{1,l},
&\text{ if }\theta =0
\end{cases}
$$
.}
\medskip

Finally we discuss the unitary group $U(n)$.\\
  The eigenvalues of matrix $M$ can be
written as:
\begin{equation}
\exp (i\theta_1),\exp(i\theta_2),\dots ,\exp (i\theta_n).d,
0\leq\theta_1,\theta_2,\dots ,\theta_n\leq 2\pi .
\end{equation}
In the rescaled coordinates $x_i=n\cdot\tfrac{\theta_i}
{2\pi},\ i=1,\dots ,n$, the $k$-point correlation functions are equal
to
\begin{align}
\begin{split}
&R_{n,k}(x_1,\dots ,x_k)=\det \biggl (\frac{\sin\pi (x_i-x_j)}{n\cdot
\sin (\pi\cdot (x_i-x_j)/n)}
\biggr )_{i,j=1,\dots k}.
\end{split}
\end{align}
In the limit $n\rightarrow\infty$ the kernel in (4.20) becomes
the sine kernel. We finish with the analoques of the last two theorems for 
$U(n)$.

\noindent{\bf Theorem 5}  {\it
Consider the normalized Haar measure on $U(n)$.  Let $\theta\in
[0,2\pi ),\ \delta_n$ be such that $0<\delta_n<2\pi -\theta -\epsilon$ for 
some $\epsilon >0$ and $n\cdot\delta_n\rightarrow +\infty$.  Denote by
$\nu_n$ the number of eigenvalues in $[\theta ,\theta +\delta_n]$.  We have
$E\nu_n=\tfrac{n}{2\pi}\cdot\delta_n $,
$$\var\ \nu_n=
\frac{1}{\pi^2}\log (n\cdot\delta_n)+\underline
{0}(1)  
$$
and the normalized random variable 
$\tfrac{\nu_n-E\nu_n}
{\sqrt{\var\ \nu_n}}$ converges in distribution to the 
normal law
$N(0,1)$
.}
\medskip

\noindent{\bf Theorem 6}  {\it
Let $ \delta _n >0 $ be such that $ \delta _n \to 0, \ n\delta_n \to \infty $
and $\nu_{k,n} = \# \left ( (k-1)\delta_n, k\delta_n] \right ). $
Then  a sequence of normalized random variables
$\tfrac{\nu_{k,n}-E\nu_{k,n}}
{\sqrt{\var\ \nu_{k,n}}}$ converges in distribution to the centalized gaussian
sequence $ \{ \xi_k \}_{k=1}^{\infty} \}$ with the covariance function
$$ E \xi_k \xi_l=
\delta_{k,l} - 1/2 \ \delta_{k,l+1} -1/2 \ \delta_ {k,l-1}
$$
.}
\medskip
\noindent{\it Remark 5}.
Results similar to Theorems 5,6 in the regime $\delta_n =\delta >0 $ 
have been also established by K.Wieand
$([W])$.

\noindent{\it Remark 6}.
For the results about smooth linear statistics in the Classical Compact Groups 
we refer the reader to  [DS], [Jo1], [So3].
\def\am{{\it Ann. of Math.} }
\def\ap{{\it Ann. Probab.} }
\def\temf{{\it Teor. Mat. Fiz.} }
\def\jmp{{\it J. Math. Phys.} }
\def\cmp{{\it Commun. Math. Phys.} }
\def\jsp{{\it J. Stat. Phys.} }
\def\npb{{\it Nucl. Phys. B} }
\def\arma{{\it Arch. Rational Mech. Anal.} }
\def\prl{{\it Phys. Rev. Lett.} }

\end{document}